# Impurity States and the Absence of Quasiparticle Localization in Disordered $D$-Wave Superconductors

A.V.Balatsky[a] and M.I.Salkola[b]

*Theoretical Division, Los Alamos National Laboratory, Los Alamos, New Mexico 87545*

(January 15, 1996)

The absence of localization of impurity-induced low-energy quasiparticle states in a two-dimensional $d$-wave superconductor is argued for any amount of disorder in the limit of unitary scatterers. This surprising result follows from the fact that a unitary impurity produces a marginally-bound state at zero energy which decays as a power-law along the nodes of the $d$-wave energy gap. Consequently, for finite density of impurities, the impurity-induced states are coupled by long-range overlaps yielding extended quasiparticle states below a characteristic energy scale $\omega_c$. Simple scaling arguments suggest that $\omega_c \propto e^{-c/n_{\rm imp}}$, where $n_{\rm imp}$ is the impurity density and $c$ is a positive constant.

When impurity scattering violates the symmetry of the superconducting condensate, the superconducting energy gap is depleted and ultimately destroyed by impurities. This happens, for example, in $s$-wave superconductors with magnetic impurities [1,2]. Some of the recent experiments support $d$-wave symmetry as a pairing channel in the cuprates [3–5], predicted by the spin-fluctuation theory [6–8] as well as by the Hubbard model [9]. For this state, even scalar impurities act as pair-breakers, producing a finite lifetime for quasiparticles near the nodes in the gap, and a finite density of states at low energies [10]. As a result, the measured low temperature properties of, for example, $YBa_2Cu_3O_7$ [3], $Bi_2Sr_2CaCu_2O_8$ [11], and $La_{1.86}Sr_{0.14}CuO_4$ [12] display a remarkable sensitivity to the presence of impurities.

Recently, the role of imperfections in $d$-wave superconductors has been considered by Lee [13], Hirschfeld and Goldenfeld [14], and subsequently by others [8,15,16]. Qualitatively these low-energy quasiparticle states behave as a disordered Fermi liquid with a renormalized density of states. Impurity scattering in a disordered $s$-wave superconductor can lead to localization of low-energy states, as was first pointed out by Ma and Lee [17] and Maekawa and Fukuyama [18]. The same effect should also, in principle, take place in disordered $d$-wave superconductors [13,16]. Lee showed that in two-dimensional (2D) disordered $d$-wave superconductors these states are subject to weak-localization corrections and thus are localized. Localization of quasiparticles was advocated as a way to recover some $s$-wave features in a $d$-wave system. Here, we argue that the conclusion regarding localization found in Refs. [13,16] is based on a simplified picture which does not take into account the long-range nature of hopping between impurity states. It is this long-range hopping which makes our approach different and which is lost upon trivial averaging over impurity positions.

In obtaining the above results, impurity-averaged quantities were expressed in terms of ensemble-averaged single-particle Green's functions which were calculated within the coherent-potential approximation (CPA) [19]. Note that, in the CPA, the quasiparticle self-energy $\Sigma(\omega) = n_{\rm imp} T(\omega)$ is approximated by a single impurity $T$-matrix; $n_{\rm imp}$ is the impurity density. Spatial anisotropy of impurity states, which is important for a $d$-wave superconductor, is lost upon averaging. Moreover, the CPA is not adequate for studying fine details of impurity bands [19].

In this Letter, we consider the problem of localization of quasiparticle states at low energies, starting from a single-impurity solution as a basis [20]. Our main result is the conclusion that the weak-localization theory does not apply to localization of impurity-induced quasiparticle states in 2D $d$-wave superconductors because of the long-range interactions between these states [21]. We argue that, in order to address this question, one has to implement an approach where the single-impurity problem is solved first. This approach was used by Shiba and Yu to study the effect of magnetic impurities in an $s$-wave superconductor [22,23]. They showed that a quasiparticle bound state is formed in the energy gap as a result of multiple scattering. A similar consideration of scalar impurities in a $p$-wave superconductor was done by Buchholtz and Zwicknagl and by Stamp [24]. These mid-gap states eventually form an impurity band in conventional superconductors. Notably, scalar unitary impurities create largely analogous impurity states in $d$-wave superconductors [20].

Below, we first argue that, due to the long-range overlaps between impurity states, disorder scattering *does not localize* quasiparticle states at $\omega = 0$ and in its vicinity. Along the directions of the vanishing energy gap (the diagonals of the square lattice for a $d_{x^2-y^2}$ gap function), the impurity wave functions decay as $1/r$. This in turn leads to the $1/r$ power-law overlaps between impurity states and to a novel network of strongly coupled impurities, yielding extended impurity states (see Fig. 1). From the point of view of the localization theory, our impurity problem belongs to a new class where for any amount of disorder long-range hopping delocalizes impurity states. Second, we speculate that the nature of low-energy states in a disordered $d$-wave superconductor is qualitatively different from the simple extended states

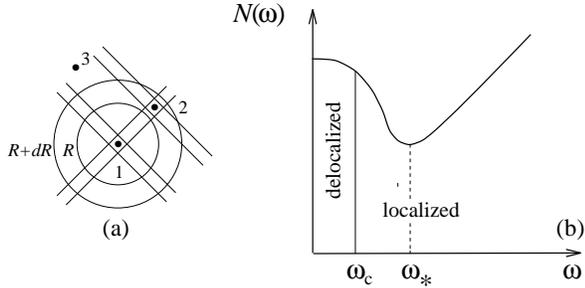

FIG. 1. (a) Three impurities are shown at 1,2, and 3. For the impurities at sites 1 and 2, the long-range $1/r$ tails of the impurity states along the nodal directions of the energy gap $\Delta(\phi)$ are depicted by two parallel lines. Strong overlap between these states leads to a resonance behavior, if the on-site energies $\epsilon_1$ and $\epsilon_2$ are close enough. The exponentially small overlap between states at sites 1 and 3 will not produce a resonance at small impurity densities, $\xi_0^2 n_{\text{imp}} \ll 1$, and are omitted reducing the original problem to a localization problem on a network of strongly coupled impurity states. The ring $(R, R+dR)$ denotes the area which specifies the probability $P_{\text{res}}(R,dR)$ [see Eq. (4)] for having another impurity resonant with the impurity at site 1. (b) The density of states $N$ of a disordered $d$-wave superconductor. Crossover from the linear dependence $N(\omega) \propto \omega$ to the impurity-dominated regime occurs at $\omega_* \propto n_{\text{imp}}^{1/2}$ for unitary scatterers [13,14]. Due to the strongly-interacting impurities, the low-energy quasiparticle states are delocalized. At higher energies, the impurity states should form a band in which weak localization should prevail [13] and there should be an "inverse" mobility edge at $\omega_c$ ($\propto e^{-c/n_{\text{imp}}}$).

predicted in any theory with averaging over random impurity ensembles.

Originally, the effect of long-range hopping on localization was considered by Anderson [26] for randomly distributed impurities in $D$ dimensions with the $V_{ij} \sim 1/r_{ij}^{D+\epsilon}$ hopping interaction (cf., RKKY-interacting spins). He showed that, for $\epsilon > 0$, there is a localization transition; for $\epsilon < 0$, states are always delocalized; and $\epsilon = 0$ represents the marginal case with power-law decaying states. Subsequently, an analogous problem of localization of phonon modes in disordered solids was considered by Levitov [27]. We will use a similar method, modified to take into account strong spatial anisotropy of impurity states.

First, consider overlaps between two impurity sites. For a unitary-scattering impurity, the bound state is well-defined and is located at zero energy. The crucial point in our approach is that, for a 2D $d$-wave superconductor, the impurity bound-state wave function is highly anisotropic and forms a cross-shape state with four tails along the diagonals of the square lattice [20]:

$$\psi_{\text{imp}}(r,\phi) \propto \sin k_F r \begin{cases} r^{-1/2} e^{-r/\xi(\phi)}, & \phi \not\simeq \frac{\pi}{4} + n\frac{\pi}{2}, \\ r^{-1}, & \phi \simeq \frac{\pi}{4} + n\frac{\pi}{2}; \end{cases} \quad (1)$$

where $\xi(\phi) \equiv \hbar v_F / |\Delta(\phi)| = \xi_0 / |\cos 2\phi|$ and $n = 0,1,2,3$ correspond to the four nodes of the gap, along which the tails are slowly decaying. We will consider, having in mind strong planar anisotropy of cuprates, a 2D $d$-wave superconductor: $\Delta(\phi) = \Delta_0 \cos 2\phi$, where $\Delta_0$ is the gap amplitude and $\phi$ is the planar angle.

In the tight-binding approximation, which is valid at low impurity densities, $\xi_0^2 n_{\text{imp}} \ll 1$, the hopping matrix element between two impurity sites $\mathbf{r}_i$ and $\mathbf{r}_j$ is given by $\hat{V}_{ij} = \hat{G}^{(0)}(\mathbf{r}_i - \mathbf{r}_j, 0)$, where $\hat{G}^{(0)}(\mathbf{r}, \omega)$ is the unperturbed Green's function in the superconducting state (for details, see Ref. [28]). We shall work in Nambu space where $\Psi_i^\dagger$ is the Nambu spinor describing a quasiparticle(hole) created in the $i$th impurity state, Eq. (1), and $\hat{V}_{ij}$ is a matrix in this space spanned by the Pauli matrices, $\hat{\tau}_\alpha$ ($\alpha = 1,2,3$) [29]. The Hamiltonian is:

$$H = \sum_{ij} \Psi_i^\dagger \hat{V}_{ij} \Psi_j + \sum_i \Psi_i^\dagger \epsilon_i \hat{\tau}_3 \Psi_i, \quad (2)$$

where the random distribution of on-site energies $\epsilon_i$ is assumed to be uniform: $P(\epsilon_i) = 1/W$, for $0 \le \epsilon_i \le W$, and zero otherwise. Since we are working with Nambu spinors, the on-site energy is a matrix too. Strictly speaking the impurity states are well defined only at $\omega = 0$; by assigning the on-site energies a distribution, these states will be hybridized with the continuum of quasiparticle states [for a clean $d$-wave superconductor, $N(\omega) \propto \omega$]. We shall ignore this small hybridization and consider only the direct overlap of the impurity-induced states.

Using Eq. (1), one can show that the overlap decays exponentially (these sites will be called weakly overlapping), *unless two impurities are connected by a radius vector* $\mathbf{r}$ *with the angle* $\phi = \pi/4 + n\pi/2$, in which case the tails of the impurity states are strongly overlapping [e.g., the impurities 1 and 2 in Fig. 1(a)]. These sites are called strongly overlapping. In this case, the hopping matrix elements decay as $1/r$:

$$\hat{V}_{ij} = -\hat{\tau}_3 V_0 (\hbar v_F / r_{ij}) \sin(k_F r_{ij} + \delta_{ij}), \quad (3)$$

where $k_F$ is the Fermi momentum, $V_0$ is the dimensionless strength of the matrix element, and $\delta_{ij}$ is a phase shift, which we will ignore hereafter. Both the $\hat{\tau}_1$ and $\hat{\tau}_3$ components of the overlap integral are present in general. However, the $\hat{\tau}_1$ component of $\hat{V}_{ij}$ is zero along the diagonals [28].

To establish delocalization, we show that for a given impurity site there exists a (large) distance $R$ within which this impurity will always find another strongly overlapping impurity. The long-range hopping will provide a strong resonance between these sites and the wave function will be delocalized [26,27]. This should be contrasted with the weak-localization theory which predicts that all quasiparticle states are localized in 2D.

We calculate the probability density for a given impurity state at site $i$ with energy $\epsilon_i$ to have a resonance with another impurity state located at a distance $R$, as shown in Fig. 1(a). For any two sites with on-site energies $\epsilon_i$ and $\epsilon_j$, the resonance will occur only for a large enough hopping, $|\hat{V}_{ij}| > |\epsilon_i - \epsilon_j|$. The $\hat{\tau}_3$ component of $\hat{V}$, $V_3(R)$, decays with the distance, whereas the phase space for possible resonances is growing with the distance and the balance for the resonance probability will depend on the power of the decay of $V_3(R)$ [26,27]. For $1/R$-decay with essentially one-dimensional motion along the tails of the wave function, we find:

$$P_{\rm res}(R, dR) = \alpha \frac{dR}{R} \ln(R/\xi_0) |\sin k_F R|, \qquad (4)$$

where $\alpha = 2\xi_0^2 n_{\rm imp} V_0 \hbar v_F / (W\xi_0)$ is a small dimensionless parameter. This small parameter allows us to ignore higher order simultaneous resonances. The average number of resonances $\int P_{\rm res}(R, dR) = N_{\rm res}$, is divergent with distance and, therefore, we are led to consider the problem of strongly overlapping random impurity states, where the tails of the impurity wave functions give rise to the network of strongly coupled impurity states [30].

To derive Eq. (4), we omit weakly overlapping sites because, for low densities $\xi_0^2 n_{\rm imp} \ll 1$, their contribution is exponentially small. The inclusion of the weakly overlapping sites will only help delocalization we are set to prove. Next, we define the angular size of the tails at large distances, as seen from the impurity site. Matching two asymptotics of the impurity wave function, Eq. (1), one finds that the exponential asymptotic matches the power law at the angle $\phi = \pi/4 \pm \delta(R)$, where

$$\delta(R) = \frac{\xi_0}{4R} \ln(R/\xi_0). \qquad (5)$$

The logarithmic factor indicates that the tails are continuously broadening at large distances. The probability density is given as $P_{\rm res}(R, dR) = \int d\phi\, P_{\rm res}(R, dR, d\phi)$ with angular dependent probability being nonzero only for $|\phi - \pi/4 + n\pi/2| \leq 2\delta(R)$ and zero elsewhere. This constraint results from the strong spatial anisotropy of the overlaps, as shown in Fig. 1(a). We find that

$$P_{\rm res}(R, dR, d\phi) = n_{\rm imp} R dR d\phi \frac{|V_3(R)|}{W}. \qquad (6)$$

Combining Eqs. (5) and (6), we finally obtain Eq. (4).

Consider the average number of resonances for a given site inside the circle of radius $R$: $N_{\rm res}(R) = \int P_{\rm res}(R, dR) \propto \alpha \ln^2 R/\xi_0$. Diverging number of resonances indicates delocalization. Another interesting result is the $\ln^2 R/\xi_0$ behaviour of the probability density. $1/R$ decay of the hopping elements along the one-dimensional tails would lead to the $\ln R$ growth of $N_{\rm res}$. However, the second logarithmic factor comes into play due to the cut-off angle $\delta(R)$ and the problem is "supercritical" instead of being simply marginal. To illustrate this point, we calculate the probability of having no resonances $P_{\rm nores}(R)$ between a given site and any other sites inside a circle of radius $R$. By drawing concentric circles $R_i$ ($i = 1, \ldots, N$), $R_{i+1} - R_i = \rho$ with $\xi_0 \ll \rho \ll R$ around the chosen impurity and calculating $P_{\rm nores}(R) = \prod_i [1 - P_{\rm res}(R_i, \rho)]$, we find *log-normal decay of no-resonance probability* instead of the expected slow power-law:

$$P_{\rm nores}(R) \propto \exp[-\mathcal{O}(1)\alpha \ln^2(R/\xi_0)]. \qquad (7)$$

One can estimate the average minimum distance between resonating sites as $R_c \propto \int R dR \exp[-\mathcal{O}(1)\alpha \ln^2 R] \propto \exp[\mathcal{O}(1)/\alpha]$, which generates an energy splitting $\omega_c = \hbar v_F/R_c$. We interpret $\omega_c$ as the characteristic energy scale of the delocalized states at $|\omega| \leq \omega_c$; see Fig. 1(b).

There is a caveat to the interpretation of the above calculation. The probability of *pairwise* resonances, Eq. (4), does not take into account the effect of closed loops, which provide *backscattering*, responsible for the localization. We believe, however, that loops are irrelevant at small $\alpha$. It is easy to show that a closed path should contain at least four sites, since at each impurity site particle can only turn by an angle $\pm n\pi/2$ in order to move along the tails. Due to this geometric constraint the probability of quartic resonances is proportional to $\alpha^4$ and does not diverge at large $R$ [28]. Therefore, at large distances pairwise resonances dominate. Based on this argument, we may conclude that inclusion of the loops should not change our basic result. In the opposite limit of no on-site disorder $W \to 0$ ($\alpha \gg 1$), the problem is in the strong coupling limit and multiple resonances are important. In this limit, the pairwise-resonance approximation is invalid and another approach is necessary. Nonetheless, it is reasonable to assume that the impurity states are delocalized again in this limit due to the strong overlaps between impurity states. However, $\omega_c$ becomes of the order of $\omega_*$ or larger and our approach breaks down because the hybridization with the continuum is a new relevant feature in the problem.

Finally, we comment on the nature of the delocalized states. In principle we are facing two possibilities: (*i*) the states are simply extended with nondecaying probability density in the whole sample. In the view of the highly nontrivial form of $P_{\rm res}(R)$, this seems unlikely, although it cannot be excluded at a moment. (*ii*) The extended impurity states are "critically" localized, *e.g.*, the impurity wave function envelope decays as $1/r^\nu$ with some index $0 \leq \nu \leq 1$. This could happen if the "dressing" of single impurity wave functions due to resonances is not sufficient to delocalize the states completely. At present we cannot distinguish between these two possibilities [31].

The results presented here are sensitive to the specific form of the energy gap close to the impurity. While the energy gap may change near the impurity, symmetry considerations [32] suggest that the node structure of the energy gap is not affected by the potential scatterers. However, purporting a more general situation, suppose that a small local imaginary $s$-wave component of magnitude $\Delta_s$ is generated in the neighborhood of the impurity. Such a component will most likely cut off the power-law tails of the wave function $\psi_{\rm imp}(\mathbf{r})$, causing it to decay as $r^{-1}e^{-r/\ell}$ along the diagonals with large $\ell = \hbar v_F/\Delta_s \gg \xi_0$. In this case, the impurity states will be localized for $\ell^2 n_{\rm imp} \ll 1$. Nonetheless, there should exist an intermediate regime, $\xi_0 \ll n_{\rm imp}^{-1/2} \ll \ell$, where the strongly overlapping impurity states along the diagonals would lead to delocalization.

*In conclusion*, we have studied formation of impurity bands in $d$-wave superconductors. We argue that long-range impurity-impurity interactions modify the usual weak-localization results and lead to extended quasiparticle states below a characteristic energy scale $\omega_c \propto e^{-c/n_{\rm imp}}$ ($c$ is a positive constant). This result suggests that unitary scatterers do not cause activated behavior of quasiparticles in quasi-2D $d$-wave superconductors at low temperatures.

We are grateful to B. Altshuler, A. Berlinsky, C. Kallin, P. Lee, M. Sigrist, D. Thouless, and S. Trugman for useful discussions. This work was supported by the U.S. Department of Energy.